\documentstyle{article}
\begin{document}
\begin{center}
{\Large {\bf Energy in Dilaton Gravity in Canonical Approach}} \\

\vspace{4mm}
M.Z.Iofa \\Nuclear Physics Institute \\Moscow State University \\Moscow
119899, Russia
\end{center}

\begin{abstract}
An expression for energy in string-inspired dilaton gravity is obtained in
canonical approach.
\end{abstract}

The problem of definition of energy in general relativity has a long history
and was repeatedly discussed in different aspects [1-6]. In standard gravity
interacting with matter, it was shown that for field configurations with
asymptotically flat metric with components sufficiently rapidly decreasing
at spacial infinity one can define a nonnegative expression which is
naturally interpreted as energy. In canonical approach, energy is defined as
the value of the standard hamiltonian taken on the shell of zero constraints
\cite{rt,lf,gt}. 
\begin{equation}
E=H|_{\{\Phi \}=0}
\end{equation}
and has a functional form of a total spacial divergence
\[
E=-\int \partial _iq^id^Dx
\]
(greek and latin indices take $D+1$ and $D$ values respectively).

In this note, following the methods of canonical approach to standard
gravity \cite{rt,lf,gt}, we discuss an expression for energy in dilaton 
gravity which appears as effective field theory for closed bosonic string 
theory. This expression can be used for calculation of mass of recently 
studied exact solutions in dilaton gravity such as black strings etc.

In $O(\alpha ^{\prime })$ order, for genus zero surfaces, effective action 
of closed bosonic string theory is \cite{gsw}

\begin{equation}
S^{\prime }=-\int d^{D+1}xe^{-\Phi }\sqrt{\left| g\right| }\left( R-(D\Phi
)^2+2D^2\Phi +\frac{H^2}{12}+\Lambda \right)   
\end{equation}
(in the following, we do not write the term $\frac{H^2}{12}$ irrelevant
for our discussion explicitly).

Let the dynamical fields be $h_{\mu\nu}$ and $\varphi$, where
$g_{\mu\nu}=\eta_{\mu\nu}+h_{\mu\nu} $ and $\Phi = \Phi^0 +\varphi$.
Here $(\eta_{\mu\nu}, \Phi^0)$ is the vacuum (flat space) solution.
Following the standard approach, we rewrite (2)
as the sum of two terms. The first term contains only first-order
derivatives, the second one is the total derivative and produces terms with
second-order derivatives. Using the relation $\sqrt{\left| g\right| }R=
\sqrt{\left| g\right| }G +\partial _\lambda(\sqrt{\left| g\right| }
w^\lambda)$ \cite{ll}, (2) is rewritten as

\begin{eqnarray}
S^{\prime } &=&-\int d^{D+1}xe^{-\Phi }\sqrt{\left| g\right| }\left(
G+\Lambda -\left( D\Phi \right) ^2+w^\lambda \partial _\lambda \Phi
+2D^2\Phi ^0+2D_\lambda \Phi D^\lambda \varphi \right) -  \\
&&-\int d^{D+1}x\partial _\lambda \left( \sqrt{\left| g\right| } e^{-\Phi
}(w^\lambda +2D^\lambda \varphi) \right).\nonumber
\end{eqnarray}
Here
\begin{equation}
w^\lambda =g^{\mu \nu }\Gamma _{\mu \nu }^\lambda -g^{\mu \lambda }\Gamma
_{\mu \nu }^\nu =g_{\mu \rho ,\nu }G^{\mu \nu \lambda \rho }  
\end{equation}
where
\[
G^{\mu \nu \lambda \rho }=-g^{\mu \nu }g^{\lambda \rho }+g^{\mu \rho
}g^{\lambda \nu }
\]
Omitting the term with the total derivative, we obtain the action with only
first-order derivatives of dynamical fields:

\begin{equation}
S=-\int d^{D+1}xe^{-\Phi }\sqrt{\left| g\right| }\left( G+\Lambda -\left(
D\Phi \right) ^2+w^\lambda \partial _\lambda \Phi +2D^2\Phi ^0+2D_\lambda
\Phi D^\lambda \varphi \right).   
\end{equation}
The fields $(g_{\mu\nu}, \Phi)$ are assumed to be asymptotic to the static
vacuum solution $(\eta_{\mu\nu}, \Phi^0)$, and dynamical fields are supposed
to decrease at spacial infinity, so that the following transformations and
formulas make sence. To make formulas more transparent (having in view, as
an example, applications to black string solutions of the gauged
$SL(2,R)\times {U(1)}^N /U(1) $ models \cite{hh,gq,bs}), 
we consider the case of configurations with $g_{0i}=0$.

Let us construct the hamiltonian for the lagrangian density (5). The
components of $w^\lambda$ are
\begin{equation}
\begin{array}{l}
w^0=\dot{h}_{ls}G^{ls00}+\bar{w}^0 \\[5mm]
w^m=\dot{h}_{0l}G^{m0l0}+\dot{h}_{ls}G^{lsm0}+\bar{w}^m
\end{array}
\end{equation}
Here the dots stand for $x^0$-derivatives, $\bar{w}^\lambda$ contain no
$x^0$-derivatives. Velocities-dependent part of the lagrangian density is

\begin{equation}
L_1 =-e^{-\Phi }a\left[- \frac 14\dot{h}_{ik}G^{iklm}\dot{h}_{lm}-
\dot{h}_{ls}g^{ls}\dot{\varphi}
+\dot{\varphi} ^2-g^{ik}\left( \partial _i\ln a-\partial _i\Phi \right)
\dot{h}_{0k}\right]. 
\end{equation}
Here 
$$a=\frac{\gamma}{\kappa},\qquad \gamma = \sqrt{|det g_{ik}|},\qquad \kappa
=\frac{1}{\sqrt{g^{00}}}.$$
The signature of the metric is $(+--\cdots )$.
The momenta are
\begin{equation}
\begin{array}{l}
p^{00}=0,\\[5mm]
p^{0k}=-g^{ik}\partial_i (ae^{-\Phi}),\\[5mm]
p^{ik}=\frac{1}{2} ae^{-\Phi}\left[(\dot{h}_{lm}g^{lm}-
2\dot{\varphi})g^{ik}+\dot{h}^{ik} \right],\\[5mm]
p_{\varphi} =ae^{-\Phi}(2\dot{\varphi}-\dot{h}_{lm}g^{lm}).
\end{array}
\end{equation}
Velocities $\dot{h}_{0i}$ are not expressed through the momenta
and yield the primary constraints
\begin{equation}
\begin{array}{l}
\Phi^{(1)0}=p^{00}=0,\\[5mm]
\Phi^{(1)k}= p^{0k}+g^{ik}\partial_i (ae^{-\Phi})=0.
\end{array}
\end{equation}
Momenta-dependent part of the hamiltonian density is
\begin{equation}
H_1 =e^{-\Phi} \frac{\kappa}{\gamma}\left[ p^{ik}g_{il}g_{km}p^{lm} +
p^{ik}g_{ik}p_{\varphi} + \frac{D-1}{4}p_{\varphi}^2 \right].
\end{equation}
Velocities-independent part of the lagrangian can be written as
\begin{eqnarray}
L_2& =&- e^{-\Phi}\kappa\gamma \left [R_D + \Lambda - D_i \Phi^0 D^i \Phi^0 +
2D_i D^i \Phi^0 + D_i \varphi D^i \varphi + (\bar{w}^l -\hat{w}^l )\partial
_l \Phi \right] + \\
&&+\partial_l (e^{-\Phi} \kappa \gamma \hat{w}^l ).\nonumber 
\end{eqnarray}
Here $\bar{w}^l$ and $\hat{w}^l$ are
\begin{equation}
\begin{array}{l}
\bar{w}^l = -\gamma\kappa \left[2g^{ln}\partial_n \ln\gamma\kappa +
\partial_n g^{ln} \right]\\[5mm]
\hat{w}^l = -\gamma \left[2g^{ln}\partial_n \ln\gamma +
\partial_n g^{ln} \right],
\end{array}
\end{equation}
$\sqrt{g}=\gamma\kappa$, $R_{D}$ is the curvature scalar constructed from spacial components of the
metric $g_{ik}$. Rearranging in (11) the terms containing derivatives of
$\kappa$ (which come from $D^i D_i \Phi^0$ and $(\bar{w}^l -\hat{w}^l )
\partial_l \Phi$) so that $\kappa$ enters either without derivatives or as
a factor in the total divergence of some expression, we get
\begin{eqnarray} 
L_2& =&-e^{-\Phi}\sqrt{|g|} \left [R_D + \Lambda -
\nabla_i \Phi \nabla^i \Phi + 2\nabla_i \nabla^i \Phi \right] +
\\ &&+\partial_l (e^{-\Phi} \sqrt{|g|}( \hat{w}^l +2g^{lk}\partial_k
\varphi)).\nonumber 
\end{eqnarray} 
Here $\nabla_i$ is covariant
derivative constructed from spacial components of the metric. Note that
the first term in (13) coincides with the $00$ component of the equation of
motion derived from (2) or (3) taken at the surface of constant $x^0$. 

Writing the second term in (13) as \begin{equation}
e^{-\Phi}\gamma(\kappa-1)(\hat{w}^l + 2\nabla^l \varphi)+
e^{-\Phi}\gamma(\hat{w}^l + 2\nabla^l \varphi) \end{equation} and provided
the combination $e^{-\Phi}\gamma(\kappa-1)(\hat{w}^l + 2\nabla^l \varphi)$
sufficiently rapidly vanishes at spacial infinity (which is true for
explicit examples of interest mentioned above), we obtain that the space
integral of the total divergence from the first term in (14) vanishes and
in (13) this term can be omitted. 

Collecting (10) and (13), we obtain the hamiltonian density
\begin{equation}
H= \kappa T^0 -\partial_l (e^{-\Phi}\gamma(\hat{w}^l + 2\nabla^l \varphi)),
\end{equation}
where
\begin{eqnarray}
T^0&= &\kappa\left \{ e^{-\Phi} \frac{1}{\gamma}\left[ p^{ik}g_{il}g_{km}p^{lm} +
p^{ik}g_{ik}p_{\varphi} + \frac{D-1}{4}p_{\varphi}^2 \right]+\right. \\
& &+\left. e^{-\Phi}\gamma \left [R_D + \Lambda - \nabla_i \Phi \nabla^i \Phi +
2\nabla_i \nabla^i \Phi  \right] \right \}  \nonumber
\end{eqnarray}
and [5]
$$ H^{(1)} =H+\lambda_\mu p^{0\mu}. $$
In general case when $g_{0i} \neq 0$, there appear terms $h_{0i}T^{i}$, and
now $T^0$ and $T^i$ depend on $g_{0i}$. As in standard gravity, the functions
$\lambda_\mu $ cannot be determined from conditions of time conservation of
the primary constraints
$$ \left\{\Phi^{\mu(1)}, H^{(1)}\right\}=0. $$
Instead, secondary constraints appear, which can be chosen as linear
combination of constraints $T^{\mu}$ (cf. \cite{lf,gt}). The set of constraints 
is defined by
the gauge (gravitational) part of the action, and by adding the dilaton
one does not change the gauge structure of the action. This follows also
from the fact that in terms of the new metric 
$\bar{g}_{\mu\nu}=g_{\mu\nu}e^{\frac{-2\Phi}{D-1}}$
the kinetic part of the action (2) assumes the standard form (however, in
this case, the metric $\bar{g}_{\mu\nu}$ is asymptotically nonflat).

Defining the energy as in (1), we finally have
\begin{equation}
E=-\int d^D x \partial_l (e^{-\Phi} \gamma( \hat{w}^l +2\nabla^l  \varphi)).
\end{equation}
In particular case of black-string solutions of gauged $SL(2,R)\times
U(1)^{N}/ U(1)$ WZW models [8-10], it can be verified that the fields decrease 
sufficiently
rapidly in directions asymptotically transverse to the string to make all the
above transformations valid. In this case, (17) should be interpreted as the
energy of the string per unit length. For $N=1$ the formula (17) coincides 
with the expression used in \cite {hh}.

It is a pleasure to thank I.V.Tyutin for helpful discussions.


\begin{thebibliography}{99}

\bibitem{adm} R. Arnowitt, S. Deser and C.W. Misner, Phys. Rev. 116 (1959)
1322, Phys. Rev. 122 (1961) 997. 

\bibitem{rt} T. Regge and C. Teitelboim, Ann. Phys. 88 (1974) 286.

\bibitem{ll} L.D. Landau and E.M. Lifshitz, Field Theory: Theoretical
Physics, v.2, Nauka, Moscow, 1973.

\bibitem{lf} L.D. Faddeev, Uspekhi Fiz. Nauk, 136 (1982) 435.

\bibitem{gt} D.M. Gitman and I.V. Tyutin, Quantization of Fields with
Constraints, Springer-Verlag, 1990.

\bibitem{bj} D. Bak, D. Cangemi and R. Jackiw, Phys. Rev. D49 (1994) 5173.

\bibitem{gsw} M.B. Green, J.H. Schwarz and E. Witten, Superstring Theory,
Cambridge, 1987.

\bibitem{hh} J.H. Horne and G.T. Horowitz, Nucl. Phys. B368 (1992) 444.

\bibitem{gq} P. Ginsparg and F. Quevedo, Nucl. Phys. B385 (1992) 527.

\bibitem{bs} I. Bars and K. Sfetsos, Phys. Rev. D48 (1993) 844.

\end{thebibliography}
\end{document}